\documentclass[12pt,preprint]{aastex}

\shorttitle{Battery Letter}

\shortauthors{Contopoulos, Kazanas \& Christodoulou}

\begin{document}

\title{The Cosmic Battery Revisited}

\author{Ioannis Contopoulos,\altaffilmark{1}
Demosthenes Kazanas,\altaffilmark{2} and
Dimitris M. Christodoulou\altaffilmark {3}}

\altaffiltext{1}{Research Center for Astronomy, Academy of Athens, 
GR-11527 Athens, Greece. \\
Email: icontop@academyofathens.gr}
\altaffiltext{2}{NASA/GSFC, Code 665, Greenbelt, MD 20771, USA. \\
Email: kazanas@milkyway.gsfc.nasa.gov}
\altaffiltext{3}{Math Methods, 54 Middlesex Turnpike, Bedford, MA 01730,
USA. \\
Email: dimitris@mathmethods.com}

\begin{abstract}
We reinvestigate the generation and accumulation of magnetic 
flux in optically thin accretion flows 
around active gravitating objects. The source of the
magnetic field is the azimuthal electric current
associated with the Poynting-Robertson drag on the electrons
of the accreting plasma. 
This current generates magnetic field loops which open up because of the
differential rotation of the flow. We show through simple numerical
simulations that what regulates the generation and
accumulation of magnetic flux near the center 
is the value of the plasma conductivity. Although the conductivity is 
usually considered to be effectively infinite for the fully ionized 
plasmas expected near the inner edge of accretion disks,
the turbulence of those plasmas may actually render them much
less conducting due to the presence of anomalous resistivity.
We have discovered that if the resistivity is sufficiently high 
throughout the turbulent disk while it is suppressed
interior to its inner edge, an interesting steady-state process is
established: accretion carries and accumulates magnetic flux 
{\em of one polarity} inside the inner edge of the disk, whereas 
magnetic diffusion  releases magnetic flux 
{\em of the opposite polarity} to large distances.
In this scenario, magnetic flux of one polarity grows and accumulates at a steady rate in the region
inside the inner edge and up to the point of equipartition when it becomes dynamically important. 
We argue that this inward growth and outward expulsion of oppositely-directed 
magnetic fields that we propose may account for the $\sim 30$~min cyclic
variability observed in the galactic microquasar GRS1915+105.

\end{abstract}


\section{Introduction}

The issue of the orgin of cosmic magnetic fields, despite much progress in the 
field, remains an open issue in  astrophysics. The problem is basically of a 
topological nature, because the magnetic field, being a solenoidal 
vector ${\bf B}=\nabla\times {\bf A}$ (${\bf A}$ is the magnetic vector 
potential), is necessarily  absent in a homogeneous and isotropic universe. 
For the same reason, the magnetic field cannot be generated by the potential 
fluid motions that result 
from the growth of scalar inhomogeneities in such a universe, as current models
contend. As such, the evolution of the magnetic field
is akin to that of vorticity ${\bf \omega = \nabla 
\times {\bf v}}$ (${\bf v}$ is the flow velocity), 
a fact that has been noticed by at least one (to our knowledge) paper 
in the literature (see Eqs.~[4] and [7] in Kulsrud et al. [1997]).
It is also of interest to note that the source of both ${\bf\omega}$ and ${\bf B}$
is proportional to  $\nabla p \times  \nabla \rho/\rho^2$ ($p$ is the pressure and
$\rho$ is the density), 
a quantity which is nonzero only in the presence of nonbarotropic fluids (Kulsrud
et al. [1997] consider as such curved shocks and photoheating). This source term, 
while generally small, is nonetheless of great importance  because in its 
absence the initial values of  {\bf $ \omega$}, {\bf B} = 0 are preserved by the evolution 
equations. It is generally referred to as the {\em Biermann battery} and leads to 
magnetic fields that are usually quite weak initially ($\sim 10^{-20}$ G) 
but thought to be subsequently amplified to the observed mean galactic values ($\sim 
10^{-6}$ G) by dynamo processes.

Some years ago, we proposed a source-term alternative to the term of the Biermann battery 
(Contopoulos \& Kazanas [1998], hereafter CK). More specifically, we 
posited that the Poynting-Robertson radiation force acting predominantly on the 
electrons of an accretion flow around an active astrophysical source may generate 
toroidal electric currents sufficiently large to support poloidal magnetic 
fields that in certain situations could approach equipartition values. The  
Poynting-Robertson source term is proportional to $\nabla\times(\ell \, {\bf\omega})$,
where
$\ell= L \, \sigma_{\rm T}/4 \pi r m_ec^3$ is the usual accreting source compactness
($L$ is the source luminosity, $\sigma_T$ is the Thompson cross-section,
$m_e$ is the electron mass, $c$ is the
speed of light, and $r$ is the distance from the center); 
as such, besides its much larger magnitude compared to the term of the Biermann battery,
this term is additionally important because it provides a hitherto unexplored, 
direct coupling  between the magnetic field and the vorticity equations of motion.

In our earlier work (CK) we concentrated our attention on the effects of the 
Poynting--Robertson source term on the evolution of the magnetic field 
within the volume of the disk alone, by solving the corresponding (induction) 
equation in that region.  Our study showed that, in a highly conducting plasma, 
currents and fields grow only for a restricted time of the order of the accretion 
time, and thus they quickly saturate to values well below equipartition (but 
still much higher than those induced by the Biermann battery). However, for  
plasma conductivity below a certain critical value (and a free inner boundary
condition that allows the accumulation of magnetic flux interior to that point), 
we found a different regime in which the magnetic field accumulated 
around the gravitating object grows at a constant rate; we then showed that in several 
astrophysically interesting systems, magnetic fields may reach even equipartition 
values within astrophysically relevant time scales.

This `unrestricted' field growth has its origin on the fact that the 
magnetic field loops, generated by the toroidal Poynting-Robertson current 
inside the disk, open up to infinity once they reach the surface of the disk 
because of its differential rotation (and the ideal MHD conditions outside the
disk). This opening up, coupled with 
sufficiently low conductivity, allows the separation of 
{\em inward advected} field of one polarity from the {\em outward diffusing} 
field of the opposite polarity. This effect of magnetic loop opening above and below 
the disk is central in our mechanism and it has been clearly described in our 
earlier paper. 
Our work was criticized by Bisnovatyi-Kogan, Lovelace \& Belinski (2002), who, 
arguing along the lines of earlier work by Bisnovatyi-Kogan \& Blinnikov (1977), 
claimed that the magnetic field grows only over an accretion time scale or 
so and saturates at values well below equipartition. In support of their view 
they provided an exact  solution of the induction equation with the 
Poynting-Robertson term included that indeed exhibited the claimed saturation 
in the ideal MHD limit, while in an Appendix to their paper they argued
that their results would not change significantly for finite plasma conductivity. 


In our opinion, the reason for the difference in the conclusions of the two 
treatments lies in the consideration of the inner boundary condition to the 
solution of essentially the same equation. In addition, the one-dimensional 
character of the model used, although it showed that unrestricted field 
growth can take place, did not help the reader to visualize the actual 
geometry of the magnetic field on the poloidal plane. We therefore decided 
to return to this problem and perform more sophisticated two-dimensional numerical 
simulations with a variety of boundary conditions which, we hope, will
convince the reader that the mechanism described in our original paper 
can produce astrophysically interesting results. In \S~2 we describe the 
setup of our simulations and in \S~3 we present the results of our calculations.
Finally, in \S~4 our results are discussed and some conclusions are drawn.

\section{Simulation Setup}

Our simulations improve on those of CK by extending the treatment
to two spatial dimensions while incorporating the essential elements
of the cosmic battery scenario. Our treatment is still linear in that
the dynamics of the accretion disk are prescribed and we solve for the 
magnetic field evolution; we expect that this simplified treatment
will serve as a guide for our upcoming detailed, resistive, numerical MHD 
simulations of the combined magnetic field and vorticity equations 
that is in preparation.

Our comptuational domain consists of two regions: a viscous, resistive, 
geometrically thick accretion disk of specified flow velocity field and an 
overlying force-free magnetosphere where ideal MHD conditions apply.
Inside the disk, the magnetic
field generated through the Poynting-Robertson azimuthal drag
on plasma electrons is dynamically insignificant (at least initially),
hence magnetic field lines follow the flow as dictated by the
time-dependent induction equation
\begin{equation}
\frac{\partial {\bf A}}{\partial t}=
\frac{L\sigma_Tv_\phi}{4\pi r^2ce}\hat{\bf \phi}
+{\bf v}\times{\bf B}
-\eta \nabla\times{\bf B} 
\label{A}
\end{equation}
(see also CK and Bisnovatyi-Kogan, Lovelace \& Belinski~2002).
Here, ${\bf A}$ is the magnetic vector potential,
${\bf B}=\nabla\times {\bf A}$ is the magnetic field,
$e$ is the electron charge,
and $\eta$ is the magnetic diffusivity.
The Poynting-Robertson radiation force on electrons is
equal to $-L\sigma_Tv_\phi/(4\pi r^2 c)\hat{\phi}$.
Henceforth, we will work in a spherical system of coordinates
($r$, $\theta$, $\phi$). We assume the disk to be Keplerian
($v_{\phi} \propto v_K$) and thick (height $h \simeq r$), i.e. 
very similar to an ADAF 
(Narayan \& Yi 1994)\footnote{The picture one may
have in mind is that of small scale turbulent
magnetic fields building-up to equipartition values due to the
magneto-rotational instability. This effectively
thermalizes the kinetic energy of the azimuthal
motion on time scales not much longer than the local
free fall time, thus leading to a disk of
height $h \simeq r$, which is precisely the ADAF geometry we
consider.}. The remaining 
unknown quantity relevant to our calculation is the radial 
velocity $v_r$ which (for $h \simeq r$) is given by
\begin{equation}
v_r = -\alpha v_K \ ,
\end{equation}
where $\alpha$ is the usual accretion disk parameter.

In an $\alpha-$disk of the type considered here,
it is natural to assume that the flow turbulence that is responsible 
for the viscosity required for accretion is also responsible for
other dissipative processes, namely the anomalous plasma resistivity 
or equivalently anomalous magnetic diffusivity (see e.g. Heitsch 
\& Zweibel 2003). 
This can be expressed through the introduction of 
the magnetic Prandtl number ${\cal P}_m$ in the expression for the 
anomalous magnetic diffusivity
\begin{equation}
\eta \sim {\cal P}_m r|v_r| 
\label{eta}
\end{equation}
\cite{RRS96}. In what follows, ${\cal P}_m$ is taken to be 
a free parameter. Once ${\bf v}$, $L$ and $\eta$ are specified,
eq.~(\ref{A}) can be directly integrated to yield the magnetic
field as a function of time and position in the interior of the
accretion flow. Note that the disk's differential 
rotation will continually wind up the field in the azimuthal direction.

The Poynting-Robertson source of poloidal electric current
extends up to the surface of the disk, and hence
magnetic field loops emerge into a magnetosphere
above and below the disk
(note that the Poynting-Robertson effect is of course at work 
everywhere, only its effectiveness drops quickly with
distance from the central luminosity source).
The physical conditions
in the magnetosphere are assumed to be very different in that 
the dominant dynamical factor is the magnetic field and not
the plasma inertia, as is the case in the disk 
interior\footnote{We assume
the simplest possible configuration, one with minimal
magnetospheric plasma loading. We also neglect the corona
region above and below the disk which is believed
to be the origin of disk wind outflows. The study of the 
plasma dynamics in that region is outside
the scope of our present investigation, and we do refer the 
interested reader to the relevant literature (e.g. Krasnopolsky, 
Li \& Blandford 2003).}. 
Under the magnetospheric ideal MHD conditions, 
electric currents develop that quickly establish
a force-free magnetic field equilibrium.
In particular, the azimuthal winding of the field described above
will generate electric currents that will flow along the
magnetic loops and will tend to open them up. It has been
shown in the solar--photosphere literature
\cite{AJJ84} that there exists a critical amount
of winding in the azimuthal direction, beyond which
each loop will quickly break into two disconnected open parts 
with their footpoints on the rotating disk.

In reality, the disk and the magnetosphere
evolve at the same time, however the evolution speed in the
magnetosphere (the Alfven speed $v_A\sim c$) is much higher than
the evolution speeds within the disk ($v_K, v_r << v_A$).
We decided not to follow the full evolution of these loops
in the magnetosphere and consider instead a sequence of
force-free MHD equilibria above and below
the disk, characterized by
\begin{equation}
(\nabla\times{\bf B})\times {\bf B}=0\ .
\label{FF}
\end{equation}
We would like to acknowledge here that force-free is
an approximation of the physical conditions in the
magnetosphere above the disk corona.
Under conditions of axisymmetry, 
eq.~(\ref{FF}) can be rewritten in spherical
coordinates as
\begin{equation}
\frac{\partial^2\Psi}{\partial r^2}
-\frac{1}{r^2\tan\theta}\frac{\partial\Psi}{\partial \theta}
+\frac{1}{r^2}\frac{\partial^2\Psi}{\partial \theta^2}=
-I\frac{{\rm d}I}{{\rm d}\Psi}\ ,
\label{FF1}
\end{equation}
We have introduced here the magnetic flux function 
$\Psi \equiv r\sin\theta A_\phi$ and the poloidal electric current
distribution $I=I(\Psi)$. The three field components are then
given by
\begin{eqnarray}
B_r & = & \frac{1}{r^2 \sin\theta}
\frac{\partial\Psi}{\partial\theta} \ , \\
B_\theta & = & -\frac{1}{r \sin\theta}
\frac{\partial\Psi}{\partial r} \ , \\
B_\phi & = & \frac{I}{r \sin\theta}\ .
\end{eqnarray}
Eq.~(\ref{FF1}) is an elliptic
equation with well-defined boundary conditions 
(the field distribution $\Psi(r,\theta)|_{\rm surface}$ 
on the surface of the disk) that allows us to determine
the magnetic field configuration in the disk's atmosphere.

As we argued above, the magnetic field loops get wound up 
by the disk's differential rotation
up to a critical amount of winding, where all loops
break into two disconnected open parts. Under
the ideal force-free MHD conditions that we have
assumed for the disk magnetosphere, this critical
configuration contains a certain distribution $I(\Psi)$ of
poloidal electric current along the magnetic field.
It is easy to check that $I(\Psi)$ is a monotonically
growing function of $\Psi$, since in that case the
Lorentz force $(\nabla\times {\bf B})\times B_\phi\hat{\phi}$
acts to open up the magnetic loop.
The exact functional form depends on the detailed 
disk-magnetosphere interaction. We decided to choose the
following simple form
\begin{equation}
I(\Psi)=-\lambda \Psi(2-\Psi/\Psi_{\rm max})\ ,
\label{I}
\end{equation}
known to us from a different problem of
magnetospheric field line opening, the problem
of the axisymmetric pulsar magnetosphere (Contopoulos, Kazanas
\& Fendt 1999).
The constant of proportionality $\lambda$ is determined by
the requirement that all magnetospheric field lines
open up to infinity because of the disk differential rotation.
$\Psi_{\rm max}$ is the maximum value of the magnetic
flux function on the surface of the disk.\footnote{
The magnetic field is generated in the disk interior,
and the total amount of magnetic flux crossing the surface
of the disk is zero, hence $\Psi(r,\theta)|_{\rm surface}$ has a maximum value 
$\Psi_{\rm max}$ at a certain distance,
and decreases again to zero farther out.}
Our choice simplifies considerably the numerical calculation
but it is obviously not unique. As we shall see, such a distribution 
of poloidal electric current does indeed lead to a largely open 
magnetospheric configuration. 

Inside the disk, the azimuthal field winding is limited by field 
diffusion as described by the poloidal component of eq.~1. One
may easily check that, in the region outside $r_{\rm in}$ where
$\eta$ is of the same order as $rv_r$, the azimuthal field is
limited to a value of the order of $v_K/v_r$ times its poloidal
value. Outside the disk, the azimuthal field winding is limited by
the opening-up of the coronal loops to a value of the same order as its
poloidal value (e.g. Aly~1984). We thus expect a small discontinuity
in the azimuthal field in a transition region across the surface of the 
disk.

The problem is now fully specified and the numerical simulations
proceed as follows:
\begin{enumerate}
\item At each timestep, we evolve the magnetic field in the 
interior of the disk through eq.~(\ref{A}). 
\item We obtain the
distribution of $\Psi(r,\theta)|_{\rm surface}$. 
\item This distribution is used as a boundary condition for the calculation
of a force-free ideal MHD magnetospheric equilibrium
via eq.~(\ref{FF1}).
\item The above sequence of steps is repeated for each subsequent timestep.
\end{enumerate}

Before we proceed with the results of our numerical integrations, we would like
to note that the magnetic-field configurations in the disk
and magnetosphere are closely coupled. It is obvious that the field
generated in the disk is the source of the magnetospheric
field. What is not obvious though it that, in the
presence of magnetic field diffusion, the magnetospheric
field also acts back on the disk's magnetic field:
the opening up of the magnetospheric
field generates field bending and 
field tension on the surface of the disk, and as a result,
the field diffuses into the interior of the disk, in the
direction that will release the magnetic tension.
This back--reaction of the magnetosphere onto the disk's magnetic
field has not been taken into account by 
Bisnovatyi-Kogan, Lovelace \& Belinski~(2002) 
in their criticism of our original paper.
This effect is a central element in our cosmic battery mechanism.

\section{Numerical Integration}

We will work in renormalized variables where distances, times, 
and magnetic fields are rescaled, respectively, to the inner radius $r_{\rm in}$, 
the dynamical time $t_{\rm in}\equiv 
r_{\rm in}/v_{K,{\rm in}}$ at the inner radius, 
and the magnetic field $B_{P-R}\equiv t_{\rm in}\times
L\sigma_T v_{K,{\rm in}}/(4\pi cer_{\rm in}^3)$ that is generated
at the inner radius within a dynamical time, viz.,
\begin{equation}
r/r_{\rm in}\rightarrow r\ , \ \
t/t_{\rm in}\rightarrow t\ , \ \
B/B_{P-R}\rightarrow B\ .
\end{equation}

As was discussed above, our simulation setup allows us to study
any accretion disk model in conjunction with
any magnetic diffusivity distribution in the disk.
The main elements of the cosmic battery
mechanism may, however, be demonstrated clearly 
by using the following simple flow pattern:
\[
v_\phi = \left\{ \begin{array}{ll}
r^3 & \mbox{$r<1$, \ $60^{\circ}<\theta<120^{\circ}$} \\
r^{-1/2} & \mbox{$r\geq1$, \ $60^{\circ}<\theta<120^{\circ}$} \\
0 & \mbox{otherwise}
\end{array}
\right. \]
\[
v_r = -0.1 v_\phi
\]
\[
v_\theta = 0.
\]
We emphasize once again that the flow pattern
dictated by the specific accretion disk model
is independent of the Poynting-Robertson generated
magnetic field, as long as the latter remains below
equipartition values. This allows us to decouple the
magnetic field evolution from the flow kinematics.

The anomalous magnetic diffusivity $\eta$ is calculated
from eq.~(\ref{eta}) at all distances. We assume that
interior to $r = 1$ the accretion disk terminates and 
the velocity profiles are different from those for $r>1$.
The behavior given above corresponds qualitatively to 
settling on a stellar surface, although one could also produce 
the appropriate behavior for accretion onto a black hole.
As the flow rotation decreases to match that of
the underlying object, the shear decreases, and
the flow becomes laminar (in the case of a black hole,
the flow again becomes laminar interior three Schwarzschild
radii because matter is in free fall there). Laminar
flow implies the decay of the MHD turbulence that led
to the anomalous disk magnetic diffusivity, and
therefore we recover perfect field-plasma
coupling, i.e. $\eta\approx 0$ inside $r<1$.
This turns out to be important in obtaining the increase in 
magnetic flux found in CK.

\begin{figure}
\includegraphics[angle=270,scale=.80]{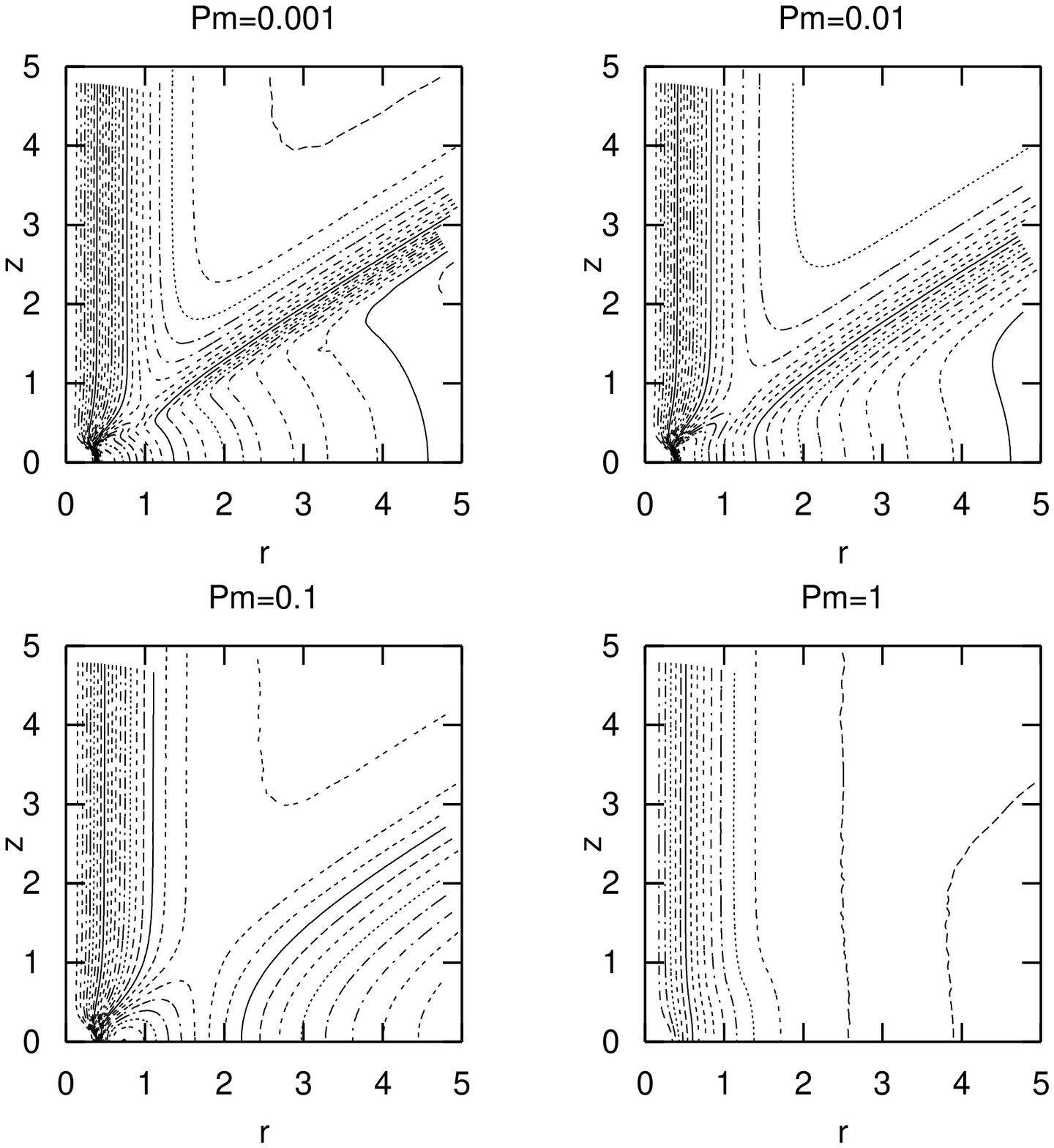}
\caption{We plot the poloidal magnetic field structure after
100 dynamical times for various values of the Prandtl number ${\cal P}_m$.
$\alpha=0.1$.
The disk extends from $60^{\circ}<\theta<120^{\circ}$, and
is surrounded by a force-free atmosphere. Magnetic field lines
that enter the atmosphere are wound up by the disk differential
rotation and open up. A poloidal electric current distribution
as defined by Eq.~\ref{I} is established in the magnetosphere.}
\label{fig1}
\end{figure}

\begin{figure}
\includegraphics[angle=270,scale=.80]{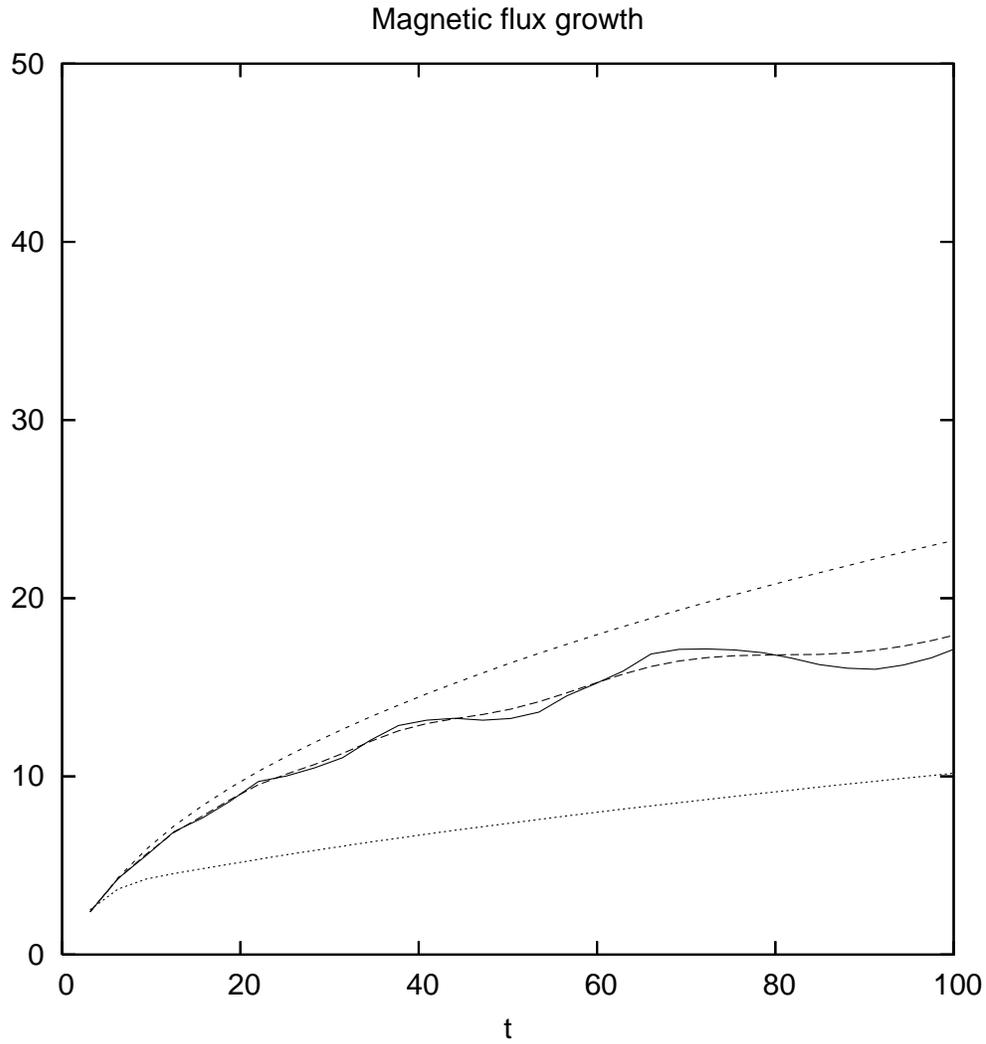}
\caption{We plot the evolution of the magnetic flux $\Psi$
contained within $r=1$ on the disk midplane for various
values of the Prandtl number (${\cal P}_m=0.001$, 0.01, 0.1, 1
for the solid, dashed, short-dashed and dotted line respectively).
In all cases, the flux accumulation saturates.}
\label{fig2}
\end{figure}

The poloidal magnetic field evolution in the disk is determined by
the $\phi$--component of eq.~(\ref{A}) written in dimensionless 
form as
\begin{equation}
\frac{\partial \Psi}{\partial t} = \frac{v_\phi}{r}
+rv_r B_\theta -{\cal P}_m rv_r \left[
\frac{\partial}{\partial r}(rB_\theta)-
\frac{\partial B_r}{\partial\theta}\right]\ ,
\end{equation}
%
%
which was integrated for 100 dynamical times. In Fig.~\ref{fig1},
we plot the resulting poloidal magnetic field configurations
for various values of the magnetic Prandtl number ${\cal P}_m$ assumed constant
throughout the disk. 
The proportionality constant in the expression for
the magnetospheric poloidal electric current (eq.~[\ref{I}]) is set equal to
$\lambda=2.5$ because
we found that, for smaller values, a large fraction of closed
magnetic field loops remains in the magnetosphere.
For small values of the magnetic Prandtl number, the magnetic
field near the inner edge of the disk
grows for about one accretion time ($\sim \alpha^{-1}$ in dimensionless
time units), and reaches an asymptotic
value of the order of $\alpha^{-1}$ (Fig.~\ref{fig2}). 
The total magnetic flux accumulated around the
center is also of the same order. 
The field O-point in the disk is
displaced from its original position around $r=1$ to near
the inner radial boundary of our computational domain.
In the high diffusivity limit, 
magnetic-field diffusion wins over accretion, 
the field O-point moves outside $r=1$, and the magnetic field 
generated by the Poynting-Robertson source eventually saturates
in this case as well.

\begin{figure}
\includegraphics[angle=270,scale=.80]{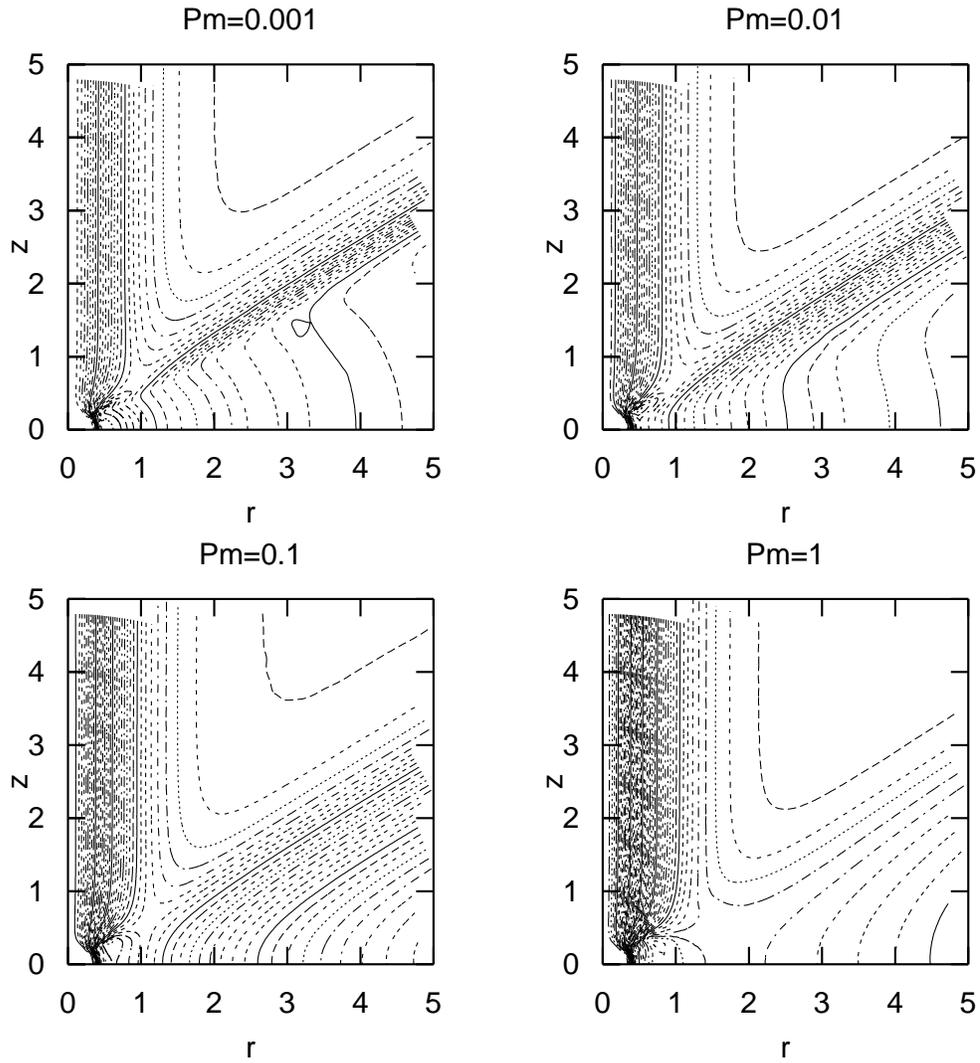}
\caption{Same as Fig.~\ref{fig1} only now we set $\eta=0$ inside
$r=1$. We see here the effect first discovered in CK,
namely that for values of the magnetic diffusivity higher
than a critical value the accumulated magnetic field grows 
without limit. }
\label{fig3}
\end{figure}

\begin{figure}
\includegraphics[angle=270,scale=.80]{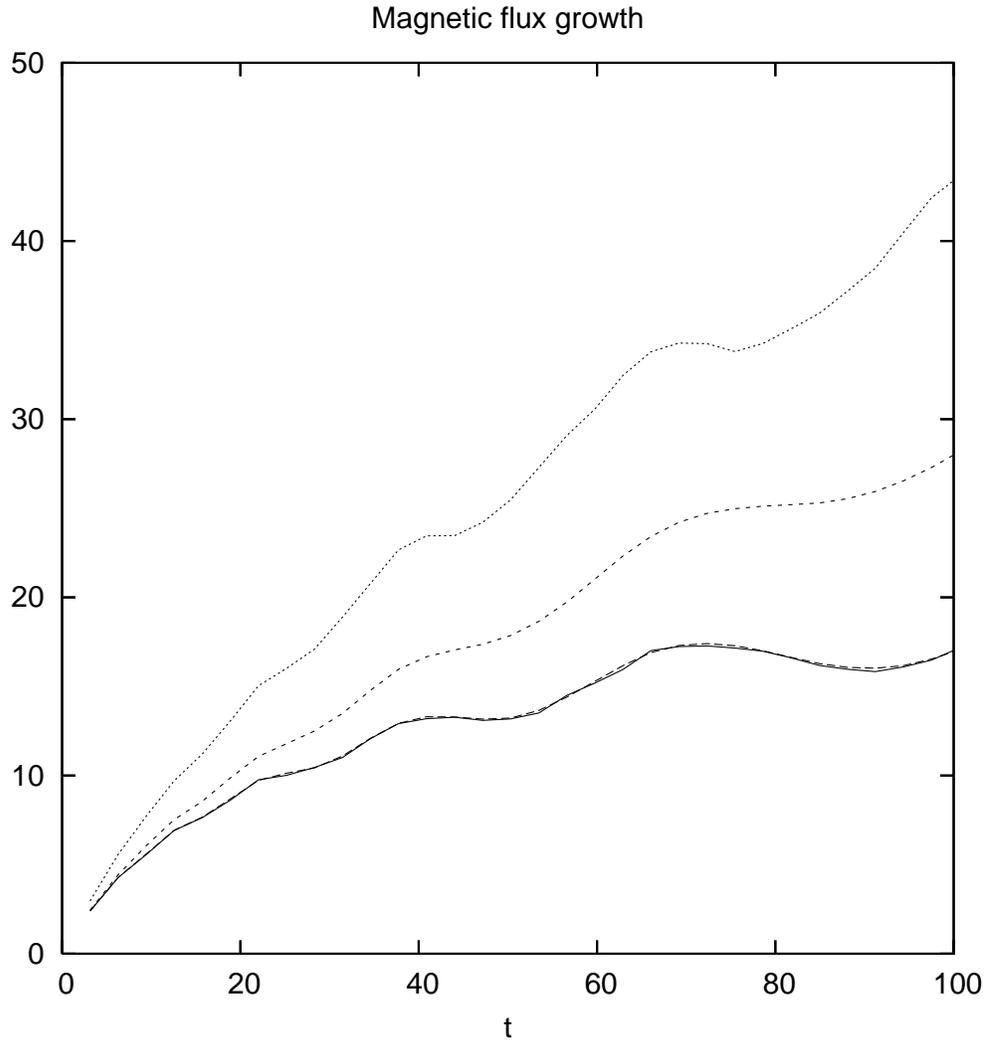}
\caption{Same as Fig.~\ref{fig2} for
values of the Prandtl number ${\cal P}_m=0.001$, 0.01, 0.1, 1
for the solid, dashed, short-dashed, and dotted
line respectively. For low values of the Prandtl number, 
the mechanism saturates
as before. Above a critical value of the Prandtl number 
${\cal P}_m\approx 1$, however, we observe unlimited steady field growth.}
\label{fig4}
\end{figure}

These results are not in contradiction with those of CK because 
CK assumed a free inner boundary for the flow that can 
only absorb the incident $B_z-$flux. To simulate this situation, 
we obtained a second set of solutions in which we set ${\cal P}_m=0$
inside $r=1$, a reasonable condition, as was discussed above. 
Our results are shown in Figs.~\ref{fig3} and \ref{fig4}.
For low values of the Prandtl number, ${\cal P}_m\ll 1$,
the field evolution is similar to the former case, namely the
field saturates within a few accretion times. However,
beyond a critical value of ${\cal P}_m\sim 1$, we obtain
a very different behavior, with new magnetic field loops centered 
around $r=1$ continuously being generated. These loops open up because of 
the disk's differential rotation and accretion 
carries inward magnetic flux of one polarity, whereas diffusion acts to remove
to large distances flux of the opposite polarity. The high 
conductivity at $r \le 1$ can then hold any magnetic flux
that enters this region. It is interesting that
the growth of $B_z$ inside $r=1$ may also
contribute to the suppression of MHD turbulence 
in that region, as seen in recent numerical simulations
(Camenzind, personal communication). 
We thus recover the effect discovered by CK, 
namely that for values of the magnetic diffusivity higher
than a critical value ${\cal P}_m\sim 1$, 
the accumulated magnetic field grows without limit.

\begin{figure}
\includegraphics[angle=270,scale=.80]{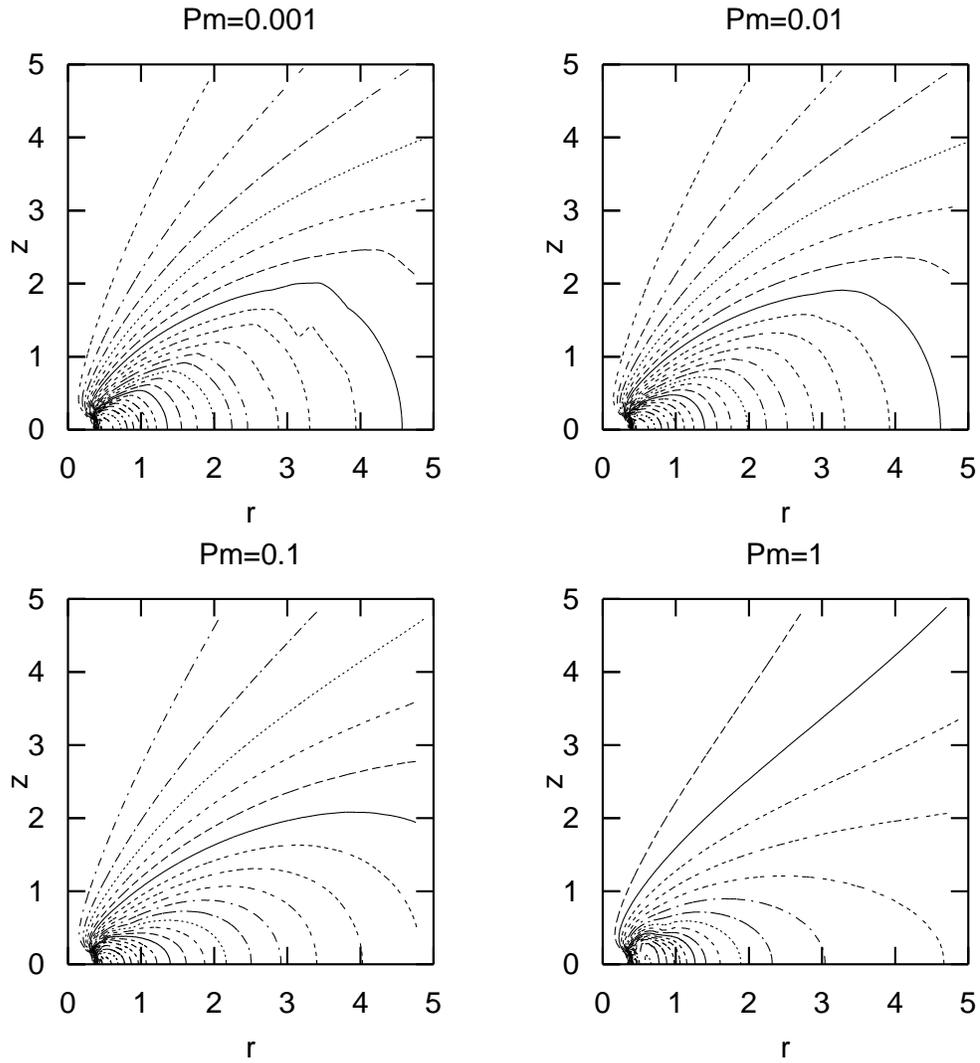}
\caption{Same as Fig.~\ref{fig3} only now we set $I(\Psi)=0$
in the magnetosphere, and the magnetic field loops do not
open up. The effect shown in Fig.~\ref{fig3}
disappears.}
\label{fig5}
\end{figure}

\begin{figure}
\includegraphics[angle=270,scale=.80]{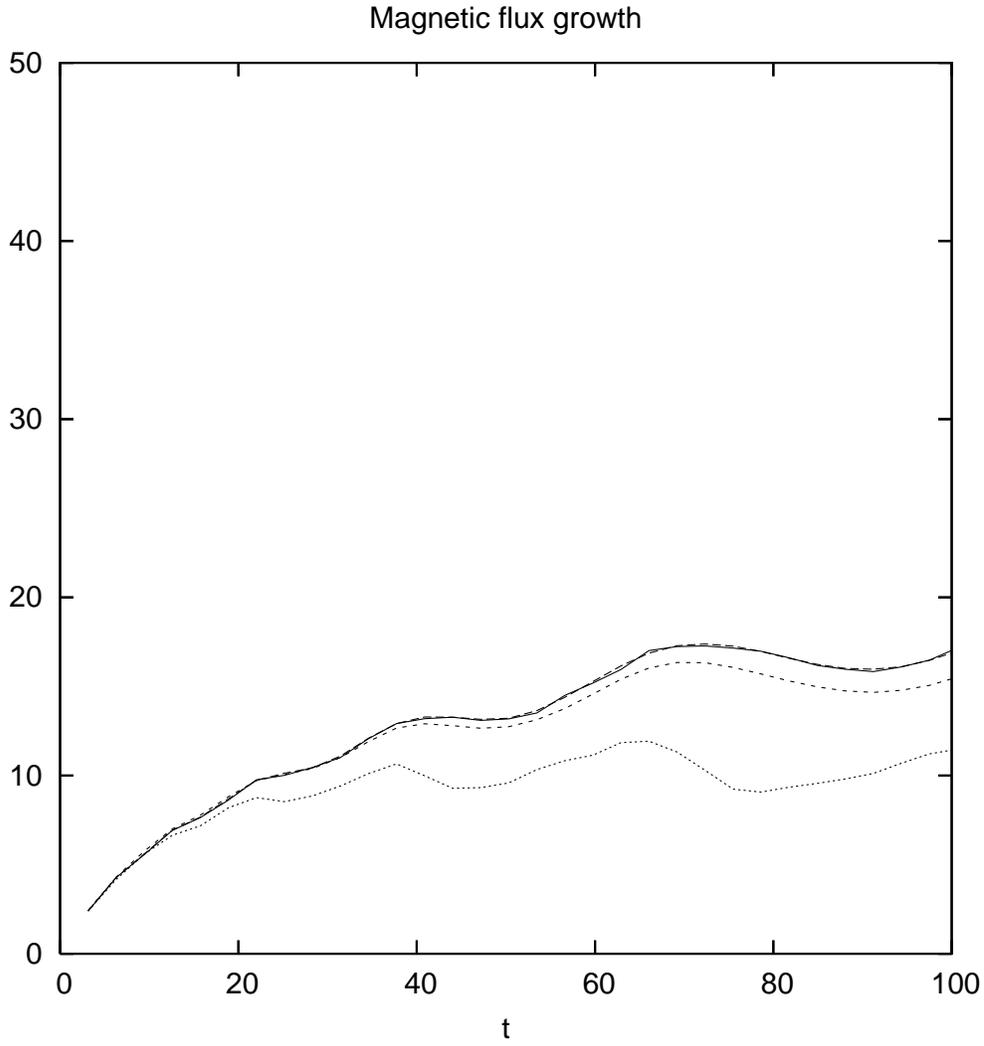}
\caption{Same as Fig.~\ref{fig4} for
values of the Prandtl number ${\cal P}_m=0.001$, 0.01, 0.1, 1
for the solid, dashed, short-dashed, and dotted
line respectively. $I(\Psi)=0$. The magnetic field again saturates.}
\label{fig6}
\end{figure}

Finally, in order to demonstrate that the opening up
of magnetic field lines is of central importance in this mechanism,
we obtained one more set of solutions in which we set $\lambda = 0$,
i.e. we ignored the magnetospheric field twisting because of the
disk's differential rotation. In that case (Figs.~\ref{fig5} and
\ref{fig6}), magnetic-field lines in the magnetosphere
remain closed and the magnetic field again saturates to a finite value
well below equipartition.

\section{Discussion and Conclusions}

We presented above  a two-dimensional analysis of the cosmic battery mechanism first
proposed in CK, paying 
particular attention to the criticism of Bisnovatyi-Kogan, Lovelace \& Belinski~(2002);
while still simplified, our 
present treatment represents a significant improvement over our earlier work, mainly
because of its two-dimensional character which is essential in capturing the structure 
and evolution of the magnetic field on the poloidal plane.

The two-dimensional models confirm our previous results. When we use appropriate
values of the plasma conductivity
and the proper magnetospheric boundary conditions, these models produce
configurations in which the magnetic flux
interior to the accretion disk's inner edge $r_{\rm in}$ increases linearly with time. 
At the same time, the present
results complement our previous investigation and delineate conditions under 
which such field growth is possible. We have thus found that, for steady magnetic-field
growth near the central object of an accretion disk, we need large diffusivity in the disk 
and a region
near the central object in which the diffusivity practically drops to zero.
The zero-diffusivity condition in the central region proves to be necessary because then
the magnetic flux that crosses into radii $r < r_{\rm in}$ cannot diffuse back out
and becomes effectively trapped near the central object
(a finite diffusivity in the central region would eventually allow for the flux to leak out to 
$r>r_{\rm in}$, saturating thus the field growth, as was argued in the Appendix of 
Bisnovatyi-Kogan, Lovelace \& Belinski~[2002]).
We believe that the zero-diffusivity condition is 
reasonable in the central region where the accretion disk does not extend 
and the associated dissipative processes are thus not operative. 

In our present simplified picture, we have ignored the transition region
between the conducting interior 
(where $\eta$ is effectively zero and the field is greatly 
amplified from a field a fraction of a Gauss to a field six or more orders of
magnitude higher) and the diffusive exterior. In reality, any small amount
of diffusivity present will
`smooth out' the abrupt field transition. Magnetic flux will continue
to accumulate as long as the thickness of the transition region remains
smaller than $r_{\rm in}$. Were the thickness of the transition region to
become of the same order as $r_{\rm in}$, the mechanism would then saturate,
and the field would diffuse outward accross $r_{\rm in}$ at the same rate
that it is brought in. If we take into account the Spitzer collisional diffusivity
estimated as $\eta_{\rm Spitzer}\sim 10^3 T_6^{1.5}\mbox{cm}^2\mbox{s}^{-1}$, where
$T_n$ is the plasma temperature in units of $10^n$~K (e.g.~Zombeck~1992),
the field growth would saturate when the ratio $B(r<r_{\rm in})/B(r>r_{\rm in})$
reaches a value of the order of $v_r r_{\rm in}/\eta_{\rm Spitzer}> 10^{12}$.
We conclude that the Spitzer resistivity does not place any practical limit to the
growth of the field interior to $r_{\rm in}$.

Finally, we conclude that this battery
mechanism could be important in accounting for the observed magnetic fields in 
several astrophysical sites as outlined in CK. Around a black hole in particular,
our mechanism generates a magnetic field of the right large--scale dipolar
topology required for the Blandford-Znajek mechanism of electromagnetic energy
extraction from a spinning black hole to work (Blandford \& Znajek~[1977]; 
Blandford~[2002]).
In this respect we would like to point the reader's attention to the 
intriguing possibility that these considerations may in fact apply to 
understanding the low frequency cyclic variability observed in the
galactic microquasar GRS 1915+105 (e.g. Pooley \& Fender [1998]; Ueda {\em et al.}
[2002]). A distinguishing characteristic of this source 
are its X-ray/IR/radio flares that repeat over time scales of $\sim 20-40$ minutes and
lead to outflows with  $\Gamma \simeq 2-3$. It is interesting to note
that if magnetic fields of order $B_z \simeq 0.1$ G generated by the mechanism 
proposed herein in one accretion time of $\sim$0.01 sec
(in a model with $M=10M_\odot$, $L=0.1 L_{\rm Eddington}$, and $\alpha=0.1$) 
were to increase linearly
with time as envisioned in this work, then after a period of $\sim$1000 seconds 
they would achieve values which could easily account for the observed flares and outflows.


DK would like to acknowledge support by INTEGRAL and HST GO grants.

\clearpage

\clearpage





\begin{thebibliography}{}
\bibitem[Aly 1984]{AJJ84}
Aly, J. J. 1984, ApJ, 283, 394
\bibitem[Bisnovatyi-Kogan, Lovelace \& Belinski 2002]{B-KLB02}
Bisnovatyi-Kogan, G. S., Lovelace, R. V. E. \& Belinski, V. A. 2002,
ApJ, 580, 380
\bibitem[Blandford 2002]{B02}
Blandford, R. D. 2002, in Proceedins of the MPA/ESO, 381
\bibitem[Blandford \& Znajek 1977]{BZ77}
Blandford, R. D. \& Znajek, R. L. 1977, MNRAS, 179, 433
\bibitem[Blinnikov \& Bisnovatyi-Kogan 1977]{BB-K77}
Blinnikov, S. I. \& Bisnovatyi-Kogan, G. S. 1977, A\& A, 59, 111
\bibitem[Contopoulos \& Kazanas 1998]{CK}
Contopoulos, I. \& Kazanas, D. 1998, ApJ, 508, 859 (CK)
\bibitem[Contopoulos, Kazanas \& Fendt (1999)]{CCKF99}
Contopoulos, I., Kazanas, D. \& Fendt, C. 1999, ApJ, 511, 351
\bibitem[Heitch \& Zweibel 2003]{HZ03}
Heitch, F. \& Zweibel E. G. 2003,  ApJ, 583, 229
\bibitem[Krasnopolsky, Li \& Blandford 2003]{KLB03}
Krasnopolsky, R., Li, Z-Y \& Blandford, R. D. 2003, ApJ, 595, 631
\bibitem[Kulsrud {\em et al.} 1997]{KCOR97}
Kulsrud, R. M., Cen, R., Ostriker, J. P. \& Ryu, D. 1997, ApJ, 480, 481
\bibitem[Narayan \& Yi 1994]{NY94}
Narayan, R. \& Yi, I. 1994, ApJ, 428L, 13
\bibitem[Pooley, Fender 1998]{PF98}
Pooley, G. G. \& Fender, R. P. 1998, in IAU Col. 164, Radio Emission from
Galactic and Extraglactic Compact Sources, eds. J. A. Zensus,
G. B. Taylor \& J. M. Wrobel, ASP Conference Series, 144
\bibitem[Reyes-Ruiz \& Stepinski 1996]{RRS96}
Reyes-Ruiz, M. \& Stepinski, T. F., 1996, ApJ, 459, 653
\bibitem[Ueda et al 2002]{U02}
Ueda, Y. {\em et al.} 2002, ApJ, 571, 918
\bibitem[Zombeck 1992]{Z92}
Zombeck, M. V. 1992, Handbook of Space Astronomy and Astrophysics 
(2nd ed.; Cambridge: Cambridge Univ. Press)
\end{thebibliography}
\end{document}